\documentclass[12pt]{article}   
\usepackage{fullpage}
\usepackage[super,sort,comma]{natbib}

\usepackage{fancyhdr}		

\usepackage{amsmath,amssymb,amsfonts}
\usepackage{algorithm}
\usepackage{algpseudocode}
\usepackage{hyperref}
\usepackage{graphicx}
\usepackage{multirow}
\usepackage{subfig}

\title{End-to-end Memory-Efficient Reconstruction for Cone Beam CT}
\author{Nikita Moriakov \and Jan-Jakob Sonke \and Jonas Teuwen \\Department of Radiation Oncology, Netherlands Cancer Institute, \\the Netherlands}

\begin{document}
\maketitle

\pagenumbering{roman}
\setcounter{page}{1}
\pagestyle{plain}

\begin{abstract}
\noindent {\bf Background:} Cone Beam CT (CBCT) plays an important role in many medical fields nowadays. Unfortunately, the potential of this imaging modality is hampered by lower image quality compared to the conventional CT, and producing accurate reconstructions remains challenging. A lot of recent research has been directed towards reconstruction methods relying on deep learning, which have shown great promise for various imaging modalities. However, practical application of deep learning to CBCT reconstruction is complicated by several issues, such as exceedingly high memory costs of deep learning methods when working with fully 3D data. Additionally, deep learning methods proposed in the literature are often trained and evaluated only on data from a specific region of interest, thus raising concerns about possible lack of generalization to other regions. \\ 
{\bf Purpose:} In this work, we aim to address these limitations and propose LIRE: a learned invertible primal-dual iterative scheme for Cone Beam CT reconstruction. \\
{\bf Methods:} LIRE is a learned invertible primal-dual iterative scheme for Cone Beam CT reconstruction, wherein we employ a U-Net architecture in each primal block and a residual CNN architecture in each dual block. Memory requirements of the network are substantially reduced while preserving its expressive power through a combination of invertible residual primal-dual blocks and patch-wise computations inside each of the blocks during both forward and backward pass. These techniques enable us to train on data with isotropic 2mm voxel spacing, clinically-relevant projection count and detector panel resolution on current hardware with 24 GB VRAM. \\
{\bf Results:} Two LIRE models for small and for large Field-of-View setting were trained and validated on a set of 260 + 22 thorax CT scans and tested using a set of 142 thorax CT scans plus an out-of-distribution dataset of 79 head \& neck CT scans. For both settings, our method surpasses the classical methods and the deep learning baselines on both test sets. On the thorax CT set, our method achieves PSNR of 33.84 $\pm$ 2.28 for the small FoV setting and 35.14 $\pm$ 2.69 for the large FoV setting; U-Net baseline achieves PSNR of 33.08 $\pm$ 1.75 and 34.29 $\pm$ 2.71 respectively. On the head \& neck CT set, our method achieves PSNR of 39.35 $\pm$ 1.75 for the small FoV setting and 41.21 $\pm$ 1.41 for the large FoV setting; U-Net baseline achieves PSNR of 33.08 $\pm$ 1.75 and 34.29 $\pm$ 2.71 respectively.
Additionally, we demonstrate that LIRE can be finetuned to reconstruct high-resolution CBCT data with the same geometry but 1mm voxel spacing and higher detector panel resolution, where it outperforms the U-Net baseline as well.\\
{\bf Conclusions:} Learned invertible primal-dual schemes with additional memory optimizations can be trained to reconstruct CBCT volumes directly from the projection data with clinically-relevant geometry and resolution. Such methods can offer better reconstruction quality and generalization compared to classical deep learning baselines. \\

\end{abstract}


\tableofcontents

\newpage

\pagenumbering{arabic}
\setcounter{page}{1}
\pagestyle{fancy}
\section{Introduction}
\label{s.intro}
Since its inception in 1960s, Computed Tomography (CT) became an import imaging modality in medicine. It is characterized by a specific acquisition and reconstruction process, in which a set of X-ray projections is first acquired for varying positions of the source and the detector and where X-rays from the source typically form a narrow fan beam. Subsequently, this projection data is processed by a reconstruction algorithm yielding either a two-dimensional slice or a three-dimensional volume. One of the variants of Computed Tomography is the Cone Beam Computed Tomography (CBCT), where X-rays from the source diverge in a wider cone-shaped beam. Both the source and the detector in CBCT typically follow circular orbits around the isocenter, and the detector is a large flat panel array. CBCT is widely used in clinic nowadays in dentistry\cite{dawood2009}, interventional radiology\cite{floridi2014} and image-guided radiation therapy\cite{jaffray2002}.

CBCT reconstruction, however, is a hard problem. Firstly, it is known\cite{maas2010, tuy1983} that the data completeness condition for exact  reconstruction of the whole volume is not satisfied for circular source/detector orbits. CBCT also inherits the imaging artifacts of classical CT such as streaking due to photon starvation in highly attenuated areas, which becomes particularly pronounced for repeated lower dose CBCT scans, and beam hardening. Furthermore, scatter-induced artifacts become more prominent due to the large panel size. These issues result in generally poor Hounsfield unit calibration, which is a limitation for applications in adaptive radiotherapy, where one would ideally use a daily CBCT scan for treatment plan adjustment without registration to a prior CT scan\cite{sonke2019}. This necessitates, along with other applications, the ongoing research on CBCT reconstruction.

In recent years, reconstruction methods based on deep learning have attracted a lot of interest in the community and demonstrated very promising results in public reconstruction challenges. For example, in the recent MRI reconstruction challenges\cite{fastmri2020, beauferris2020} deep learning methods have strongly outperformed the classical baselines. Generally speaking, any medical image reconstruction task can be viewed as an abstract inverse problem for a suitable forward operator, and different approaches have been proposed in the literature for solving such problems with deep learning\cite{schoenlieb2019}. One of the possible ways to apply deep learning to CT or CBCT reconstruction problems is to use a neural network as a learned post-processing operator for a classical reconstruction method such as filtered back-projection (FBP). This strategy was investigated in a number of publications\cite{unser2017}, where it has been demonstrated that such learned post-processing can increase the reconstruction quality. In a more recent work\cite{zhi2021} the authors propose a method for 4D CBCT reconstruction that applies a deep CNN to (gated) FBP reconstructions from phase-resolved projection data, showing image quality improvement over classical algorithms. We will only consider supervised reconstruction methods in this paper, but we would like to mention that unsupervised methods have also been developed. In Noise2Inverse approach\cite{hendriksen2020}, a form of self-supervision was used to denoise FBP reconstructions in two-dimensional tomography. In adversarial regularization approach\cite{lunz2018}, the authors proposed a learnable adversarial regularizer term, similar to a critic network in GANs, which is trained to distinguish input reconstructions obtained via classical methods from high quality ground truth reconstructions without the need for paired data. The adversarial regularizer is subsequently used to refine the input reconstructions via minimizing a combination of the data consistency loss and the adversarial regularizer loss with respect to the reconstruction tensor. For CBCT in particular, a hybrid statistical iterative CBCT reconstruction method\cite{chen2018} has been introduced which combined iterative CBCT reconstruction with a denoising/deblurring neural network trained separately on natural images. During inference, the algorithm performs iterations of a denoising/deblurring step followed by optimizing CBCT data consistency.

In the case of learned post-processing, the neural network does not have direct access to the raw data, thus it can fail to recover from some of the artifacts introduced by the classical reconstruction algorithm. A rich family of alternative methods is given by \textit{learned iterative schemes}. Such schemes are often inspired by classical iterative schemes, combining the knowledge of the forward operator and its adjoint with neural networks that complement these schemes by e.g. filtering noise in the update term. A particularly important example of such schemes for two-dimensional CT is the Learned Primal-Dual (LPD) algorithm\cite{Adler2017b} which was inspired by the Primal-Dual Hybrid Gradient method\cite{pdhg2011}. In LPD, computations are performed by \textit{primal blocks} in the image domain and by \textit{dual blocks} in the projection domain, where each block is a small residual convolutional neural network, and the blocks are connected by projection and backprojection operators, enabling end-to-end training. Such architecture allows to filter noise efficiently, since raw projection data is provided to the dual blocks. Extensions of LPD to other modalities have been proposed as well, e.g., DBToR\cite{teuwen2021} has shown good results in two-dimensional Digital Breast Tomosynthesis and XPDNet has performed very well on two-dimensional MRI reconstruction\cite{ramzi2020a}. We note, however, that there are also recent examples of learned iterative schemes for CT reconstruction that work in image domain only such as AirNet\cite{airnet}.

Unfortunately, LPD does not scale easily to a three-dimensional modality such as CBCT due to memory limitations. Indeed, for a $256 \times 256 \times 256$ float tensor a single convolution layer with $96$ features would already require 12 GB memory to perform a backpropagation. In each primal and each dual block there are 2 convolution layers with 96 filters each and one convolution layer with 5 filters. Thus, for LPD of length 8 with 720 projections, a $256 \times 256$ detector panel and $256 \times 256 \times 256$ volume, over 700 GB memory in total would be required\footnote{This estimate is obtained by observing that $(12 \textrm{ GB} \times 2 + 33 \textrm{ GB} \times 2 + 0.05 \textrm{ GB}) \times 8 = 720 \textrm{ GB}$.}. This makes it hard to train LPD on clinically relevant resolutions. Increasing complexity of the primal/dual blocks beyond the simple residual Convolutional Neural Networks would increase memory requirements even further. $\partial $U-Net, which is an alternative, simpler scheme that does not operate in the projection space, was proposed\cite{hauptmann2020}. Memory footprint of $\partial $U-Net was reduced using a multiscale approach, and reconstructions obtained by primal blocks at different scales are merged together by a U-Net. However, this method still does not allow to train using clinically relevant resolutions, and the expressive power of this scheme is reduced due to the absence of dual blocks and the conservative filter counts that were necessary to reduce memory footprint. iLPD, or invertible learned primal-dual method, has been considered\cite{rudzusika2021}, where it was shown that it substantially reduces memory requirements and allows to use longer learned iterative schemes. iLPD has been combined\cite{bajic2022} with splitting the scanning geometry in 3D helical CT setting. More precisely, projection data is split into chunks along the `time axis' and for each such chunk of the projection data the corresponding chunk of the scanned volume is taken, then the resulting pair can be used to train the neural network. However, these approaches do not scale well in a situation when explicit geometry splitting is impossible as is the case for CBCT with circular trajectory, where any single projection image can in principle contain information from the whole volume. In this situation, each invidual primal-dual block must process the whole volume or the whole stack of projection images at once, making it impossible to use complex architectures or sufficient filter counts inside these blocks due to the GPU memory limitations.

In this work we propose a memory-efficient learned primal-dual scheme that circumvents these limitations, does not require any explicit splitting of the geometry and works in 3D CBCT setting for clinically relevant resolution and projection count. The key results and novelties of this work are:
\begin{itemize}
    \item We develop LIRE, a practical framework for deep learning-based CBCT reconstruction with clinically-relevant resolution and projection count using a learned iterative scheme that can be trained end-to-end on current GPUs with 24 GB VRAM. Our framework is comprised of a learned primal-dual iterative scheme with invertible residual primal-dual blocks, and a particular set of essential memory optimization techniques that are embedded in the algorithm. To compute the gradients, only the final latent vectors and the sequence of reconstructions returned by the algorithm are needed. This allows to use longer schemes, but on its own does not allow to use complex primal-dual blocks with 3D data due to memory limitations. We additionally rely on the local nature of Convolutional Neural Networks (U-net included) and perform patch-wise computations inside the primal-dual blocks during both training and evaluation. During backpropagation, weight gradients received from different patches are summed, giving correct global gradients for network weights. Thanks to these novelties, we are able to use 3D U-nets with high filter count inside the primal blocks. Conceptually, our framework allows to use U-nets in both primal and dual blocks, which can be important for scatter correction but also when applying this framework to other modalities such as MRI. We provide the network with an auxiliary scalar tensor which has the same shape as the reconstructed volume. In this tensor, intensity of a given voxel contains information about the percentage of projections for which the corresponding spatial location is visible. The network is trained to reconstruct the intensities of all voxels which are visible in at least one projection, which results in a larger field-of-view than FBP.
    \item Two models are trained from scratch for clinical CBCT geometries with small and large field-of-view respectively, where the large field-of-view is accomplished via an offset of the detector panel. We train the models on thorax CT scans with $256^3$ voxels ($2$ mm voxel pitch), using a $256 \times 256$ detector panel ($1.6$ mm pixel pitch) and either $400$ or $720$ projections.
    \item We demonstrate superiority of our method to analytical, iterative and deep learning baselines on the test set of thorax CT scans for both field-of-view settings. Since learned iterative schemes are memory and compute intensive, we perform toy experiments to compare LIRE against alternative learned iterative reconstruction schemes such as LPD, iLPD and $\partial$U-Net. 
    \item We demonstrate better out-of-distribution generalization of our method compared to a deep learning baseline for both geometries on a test data set of head \& neck CT scans, where our method improves upon analytical and iterative baselines as well.
    \item We show additionally that using NVIDIA A100 Tensor Core GPUs with 80 GB VRAM our model can be easily fine-tuned on thorax CT scans with $512^3$ voxels at a native $1$ mm voxel pitch, $512 \times 512$ detector panel and $720$ projection images. We compare the fine-tuned model with a deep learning baseline, observing superiority of our method.
\end{itemize}

\section{Methods}
\label{s.matmethods}

\subsection{Tomography and inverse problems}
\label{ss.tomo}
CBCT reconstruction can be viewed as an inverse problem. Let $x: z \mapsto x(z)$ be a function specifying the attenuation coefficient for every point $z \in \Omega_X$ in the spatial domain $\Omega_X \subset \mathbb R^3$. The circular source rotation orbit is parametrized as a curve $\gamma: [0,1] \to \mathbb R^3$. Detector position and orientation are specified as a family of planes $\Omega_Y: t \mapsto \Omega_Y(t)$ for $t \in [0,1]$, where each such plane is canonically identified with $\mathbb R^2$. The line going from the source position $\gamma(t)$ at time step $t \in [0, 1]$ to the detector element $u \in \Omega_Y(t)$ is denoted by $L_{t, u}$. The \textit{cone-beam transform operator}, or simply the \textit{projection operator}, is then defined as
\begin{equation}
\mathcal P(x)(t, u) = \int_{L_{t, u}} x(z) dz, 
\end{equation}
therefore, $\mathcal P$ is a linear operator mapping functions defined on $\Omega_X$ to functions defined on $[0,1] \times \mathbb R^2$. Hermitian\footnote{For suitably defined $L^2$ function spaces.} adjoint $\mathcal P^*$ of $\mathcal P$ is called the \textit{backprojection operator}.

Noisy CBCT data acquisition process can then be modeled as
\begin{equation}\label{eq.noisemodel}
y = \text{\texttt{Poisson}}(I_0 \cdot e^{-\mathcal P x}),
\end{equation}
where $I_0$ is the unattenuated X-ray photon count. The inverse problem of CBCT reconstruction is then to determine the tissue attenuation coefficients $x$ given the noisy projection data $y$.

The inverse problem in \eqref{eq.noisemodel} will be approached via computing a Bayes estimator, which is parametrized by a neural network. Let $\mathrm X \times \mathrm Y$ be the probability space of pairs of tomographic volumes $x \in \mathrm X$ and their corresponding projection images $y \in \mathrm Y$ with the joint probability density function $\pi(x, y)$. Let $\hat x: \mathrm Y \to \mathrm X$ be an estimator and let $L: \mathrm X \times \mathrm X \to \mathbb R$ be a fixed cost function. The goal for the Bayes estimator $\hat x_{\textrm{Bayes}}$ in general is to minimize the expected cost
\begin{equation}\label{eq.bayesloss}
L(\hat x) = \mathbb E_{(x, y) \sim \pi} \ L(x, \hat x(y))
\end{equation}
over all estimators $\hat x$. In this work, a sum of mean absolute error and a Structural Similarity loss will play the role of the cost function $L$, and the optimal estimator will be chosen from a certain class of neural networks. Minimization of the cost in \eqref{eq.bayesloss} with respect to the parameters of the network will be carried out via minibatch stochastic gradient descent. That is, minibatches of tomographic volumes and their corresponding projections will be repeatedly sampled from the training set, the corresponding reconstructions will be computed, and the cost in \eqref{eq.bayesloss} will be optimized via stochastic gradient descent on the network parameters. We refer to Section 5.1.2\cite{Kaipio2005} for more information on Bayes estimators.

\subsection{Data}
\label{ss.data}
In this work we simulate two common clinical acquisition geometries for a  Linac-integrated CBCT scanner from Elekta\cite{letourneau2005}: a small field-of-view setting and a medium field-of-view setting, which will refer to as `small FOV setting' and `large FOV setting'. For both settings, the source-isocenter distance is $1000$ mm and the isocenter-detector plane distance is $536$ mm. For the small FOV setting, the source-isocenter ray passes through the center of the detector, while for the large FOV setting the detector is offset by $115$ mm to the side in the direction of rotation. Square detector panel with a side of $409.6$ mm and $256 \times 256$ pixel array was used for the main experiments, while for the additional high resolution study we swtiched to $512 \times 512$ pixel array. The projection counts were $400$ and $720$ for the small FOV and the large FOV setting respectively for the main experiments. For the toy experiment we used a reduced projection count of $64$, a $128 \times 128$ pixel array and a large FOV setting. The source moves over a $200$ degrees arc for the small FOV setting, and for the large FOV setting the source moves over a full circle.

To train and evaluate our models, we collected two diagnostic CT datasets from the institutional archive: a dataset of 424 thorax CT scans with isotropic spacing of $1$ mm, and a dataset of 79 head \& neck CT scans with anisotropic spacing of between $0.9$ mm and $1.0$ mm for axial plane and between $1.0$ mm and $1.6$ mm for the perpendicular direction. Both datasets had axial slice of $512 \times 512$ voxels. For the toy experiment, all data was downsampled by a factor of $4$, resulting in volumes with $128^3$ voxels. For the main experiments, all data was downsampled by a factor of $2$, resulting in volumes with $256^3$ voxels. For the additional full resolution study, we did not apply any resampling. Study approval was granted by the IRB of our institute, IRBd20-008.

The thorax CT dataset was used to train, validate and test the models, while the additional head \& neck dataset was used exclusively for testing the models on out-of-distribution data. The thorax CT dataset was partitioned into a training set of 260 scans, a validation set of 22 scans and a test set of 142 scans. 

To simulate noisy projection data for the CT scans, Hounsfield units were converted into attenuation coefficients using $\mu = 0.2 \ \textrm{cm}^{-1}$ as the water linear attenuation coefficient. Attenuated projection data was corrupted by Poisson noise with $I_0 = 30000$ photons in \eqref{eq.noisemodel}.

\subsection{Baseline methods}
\label{ss.baselines}
First of all, we performed a toy experiment, where a version of LIRE with reduced filter count, denoted as LIRE-32, was compared against other recent learned iterative schemes such as a memory-efficient algorithm based on penalized weight least squares reconstruction method that does not need end-to-end training\cite{wu2019} which we denote as DL-PWLS, LPD\cite{Adler2017b}, two versions of $\partial$U-Net\cite{hauptmann2020}, an iLPD\cite{rudzusika2021} and an even lighter version of LIRE-32, denoted as LIRE-32(L), where the primal blocks are simplified and share the same architecture as the primal blocks in LPD. LPD and LIRE-32(L) utilise primal/dual blocks with 3 convolutional layers having (32, 32, 5) filters in case of LPD and (32, 32, 4) filters in case of LIRE-32(L). LIRE-32 uses dual blocks with a similar stack of 3 convolutional layers with (32, 32, 4) filters, while primal blocks are all U-Nets\footnote{See Section \ref{ss.arch} for full description of LIRE architecture.} with 32 filters in the top layer of the U-Net and 64 filters in the bottom layer of the U-Net. Our implementation of $\partial$U-Net\cite{hauptmann2020} relies on the open-source implementation\footnote{Adapted to 3D and our projector/backprojector code from \url{https://github.com/asHauptmann/multiscale}} from the author; however, the filter counts were increased from 12 to 16 in $\partial$U-Net-16 and to 32 in $\partial$U-Net-32 in the primal blocks of the scheme; the base filter count of the U-Net block of $\partial$U-Net was also increased from 12 to 16 in $\partial$U-Net-16 and to 32 in $\partial$U-Net-32. For iLPD, we used the official implementation\footnote{Adapted to our projector/backprojector code from \url{https://github.com/JevgenijaAksjonova/invertible_learned_primal_dual}}, and considered a version of iLPD with 32 filters in primal/dual layers and length of 8. For DL-PWLS, we used 8 blocks of U-Nets with a single downsampling layer and 32 base filters, similar to LIRE-32, in order to obtain a comparable parameter count. Information about the field of view was not provided to any of the models in the toy experiment. We used a variant of the loss from Alg.\ \ref{alg:liretrain} by setting the parameters $\alpha_1 = 0.0, \alpha_2 = 0.01$\footnote{I.e., we do not include SSIM term and have very small weight for the reconstruction error outside the full field of view.} for training all the the toy baselines.

For the main experiments in this work we considered the following baselines: FBP\cite{feldkamp84}, PDHG\cite{pdhg2011} with TV regularisation, U-net\cite{cicek2016} and Uformer\cite{wang2022} with FBP input. 

We used ODL\cite{odl2017} implementation of FBP and PDHG. We chose Hann filter with $0.9$ frequency modulation for the FBP baseline by testing different combinations of filter and frequency on the training set. Parker weighting\cite{wesarg2002} was used for FBP reconstruction with small FOV. For FBP reconstruction with large FOV setting it is important to take into account the fact that the central cylindrical region of the volume is measured by twice as many projections as the rest of the FOV, this results in a reconstruction artifact in the form of a bright ring around the center of axial volume slices. To reduce this effect, one solution is to smoothly reduce intensity of projections in the detector region which captures twice as much data as we go from the detector center to the detector edge. We do so by multiplying all projections by the following weighting factor\cite{microct, wang2002} after the FBP filtering and before the backprojection:
\begin{equation*}
\omega(s) = \begin{cases} 
    1 & -\Delta \leq s \leq -\Theta \\
    \frac 1 2 \left(  - \sin \left( \frac{\pi \arctan (s / D)}{2 \arctan(\Theta/ D)} \right) + 1\right) & -\Theta \leq s \leq \Theta \\
    0 & \Theta \leq s \leq \Delta 
   \end{cases}
\end{equation*}
In this formula, $s$ is the signed distance between detector a pixel and the projection of the rotation axis onto the detector plane, which is taken with the `minus' sign if we are closer to the detector center than to the edge and with the `plus' sign otherwise, $D$ is the size of the detector, and $\Theta = 0.289 D$ is a parameter which we chose experimentally for our geometry to obtain uniform reconstructions without the ring artifacts.

For the PDHG baseline, we used $600$ iterations with $0.25$ weight for the TV regularization term. The parameters of PDHG were obtained via tuning on the train set as well.

Finally, as the main deep learning baseline we implemented a 3D U-net for post-processing the FBP output and compared it to a two-dimensional post-processing baseline using a more recent Uformer model\cite{wang2022}, variants of which have lately been used for CBCT post-processing in the literature\cite{yang2023, chen2022}. We used U-net with $3$ downsampling layers, PReLU activations, valid convolutions and $64$ base filters, similar to the original 3D U-net\cite{cicek2016}, but without Instance or Batch normalization layers. As input for the U-net and the Uformer, we provided the FBP reconstruction and the field-of-view tensor $V_f$ defined later in Section \ref{ss.liretrain}. Uformer-S variant from the official implementation was chosen, where the transformer channel count was increased from the default value of 32 to 64. Two U-net's and two Uformer's, one for small FOV and one for large FOV, were trained for the main experiments on downsampled data using the same augmentation strategy as LIRE and the same loss function as LIRE (see Alg.\ \ref{alg:liretrain}), except for the reconstruction loss over the partial field of view region which we omit by setting $\alpha_2 = 0.0$, since FBP quality is very poor in this region and we do not evaluate the baselines in this region either. U-net's were trained to reconstruct $128 \times 128 \times 128$ patches due to memory limitations, while Uformer was reconstructing complete $256 \times 256$ axial slices. Adam optimizer\cite{kingma2014} was employed with an initial learning rate of $0.0001$ and a plateau scheduler with linear warm-up and 10 epoch patience. The best-performing model on the validation set was used for testing. A separate U-net with similar architecture was trained for the full resolution study on high-resolution data; $L^1$ reconstruction loss restricted to the full field of view was minimized in this experiment. One NVIDIA Quadro RTX 8000 was used for training the U-net's and the Uformer's.

\subsection{LIRE architecture and implementation}

\label{ss.arch}
\begin{algorithm}
\centering
\begin{algorithmic}[1]
    \Procedure{\texttt{reconstruct}}{$y, \mathcal P, \mathcal P^*, \theta, V$}
        \State $x_0 \gets \mathcal P^*(y)$ \Comment{Normalized backprojection init}
        \State $I \gets []$ \Comment{Initialize output list}
        \State $f \gets x_0^{\otimes 8} \in X^{8}$\Comment{Initialize primal vector}
        \State $h \gets y^{\otimes 8} \in U^{8}$\Comment{Initialize dual vector}
        \For{$i \gets 1, \dots, 8$}
        \State $d_1, d_2 \gets \text{\texttt{Splt}}(h)$ \Comment{Split dual channels}
        \State $p_1, p_2 \gets \text{\texttt{Splt}}(f)$ \Comment{Split prime channels}
        \State $p_{\text{op}} \gets \mathcal P([p_2, x_{i-1}]^{\oplus})$ \Comment{Project $p_2$ and $x_{i-1}$}
        \State $d_2 \gets d_2 + \Gamma_{\theta_i^d}([p_{\text{op}}, d_1, y]^{\oplus})$ \Comment{Upd. $d_2$}
        \State $b_{\text{op}} \gets \mathcal P^*(d_2)$ \Comment{Backproject $d_2$}
        \State $\text{\textit{LW}} \gets \mathcal P^* (\mathcal P (x_{i-1}) - y)$ \Comment{Landweber term}
        \State $p_2 \gets p_2 + \Lambda_{\theta_i^p}([b_{\text{op}}, p_1, x_{i-1}, \text{\textit{LW}}, V]^{\oplus})$ \Comment{Upd. $p_2$}
        \State $h \gets [d_1, d_2]^{\oplus}$ \Comment{Combine new dual}
        \State $f \gets [p_1, p_2]^{\oplus}$ \Comment{Combine new primal}
        \State $x_i \gets x_{i-1} +  \text{\texttt{Conv3d}}(f, \theta_i^o)$ \Comment{Update $x_{i-1}$}
        \State $I \gets I + [x_i]$ \Comment{Append $x_i$ to output list}
        \State $h \gets \text{\texttt{Perm}}(h, {\theta_i^m})$ \Comment{Permute dual channels w. $\theta_i^m$}
        \State $f \gets \text{\texttt{Perm}}(f, {\theta_i^m})$ \Comment{Permute prim. channels w. $\theta_i^m$}
        \EndFor
        \State \textbf{return} $I$ 
    \EndProcedure
\end{algorithmic}
\caption{LIRE.}
\label{alg:liremain}
\end{algorithm}
\begin{algorithm}
\centering
\begin{algorithmic}[1]
    \Procedure{\texttt{loss}}{$x, y, V_f, V_s$}
        \State $L_1 \gets \| x - y \|_{V_f, 1}$ \Comment{$L^1$ loss in full FOV}
        \State $L_2 \gets \| x - y \|_{V_s, 1}$ \Comment{$L^1$ loss in part. FOV}
        \State $S_1 \gets 1.0 - \text{\texttt{SSIM}}_{V_f}(x, y)$ \Comment{1-SSIM, full FOV}
        \State $S_2 \gets 1.0 - \text{\texttt{SSIM}}_{V_s}(x, y)$ \Comment{1-SSIM, part. FOV}
        \State \textbf{return} $L_1 + \alpha_1 S_1 + \alpha_2 (L_2 + \alpha_1 S_2)$
    \EndProcedure
    \For{$j \gets  1, \dots, N_{\text{iter}}$}
        \State $x \sim \mathcal{D}_{\text{train}}$\Comment{Sample train volume}
        \State $\Delta \sim \mathcal N(0, 100) \in \mathbb R^3$ \Comment{Sample offset w.r.t. scan center}
        \State $\delta \gets x.\text{\texttt{spacing}}$ \Comment{Get spacing of volume $x$}
        \State $\mathcal P, \mathcal P^* \gets \mathcal P_{\Delta, \delta}, \mathcal P_{\Delta, \delta}^*$ \Comment{Define projector, backprojector}
        \State $\overline{\mathcal P}, \overline{\mathcal P}^* \gets \mathcal P / \| \mathcal P\|, \mathcal P^*  /  \| \mathcal P\|$ \Comment{Normalize operators}
        \State $y \gets \text{\texttt{Poisson}}(I_0 \cdot e^{-\mathcal P(x)})$ \Comment{Noisy projections}
        \State $\overline y \gets -\text{\texttt{ln}}(y) / \| \mathcal P\| $\Comment{Normalized log-transform}
        \State $V_f \gets \text{\texttt{FullFOV}}(\mathcal P)$ \Comment{Compute full FOV}
        \State $V_p \gets \text{\texttt{PartialFOV}}(\mathcal P)$ \Comment{Compute partial FOV}
        \State $V_s \gets V_p \setminus V_f$ \Comment{Incomplete FOV mask}
        \State $I \gets \text{\texttt{RECONSTRUCT}}(\overline y, \overline{\mathcal P}, \overline{\mathcal P}^*, \theta, V_f)$ \Comment{Reconstruct}
        \State $\text{loss} \gets 0$ \Comment{Initialize loss tensor}
        \For{$z \gets  I[1], \dots, I[8]$} \Comment{Loop over iterates}
        \State $\text{loss} \gets \text{loss} + \text{\texttt{LOSS}}(x, z, V_f, V_s)$ \Comment{Increment loss}
        \EndFor
        \State $\text{compute gradients of loss  w.r.t. $\theta$, update $\theta$}$
    \EndFor
\end{algorithmic}
\caption{Training of LIRE.}
\label{alg:liretrain}
\end{algorithm}



\begin{figure*}[!ht]
    \centering
    \includegraphics[width=0.7\linewidth]{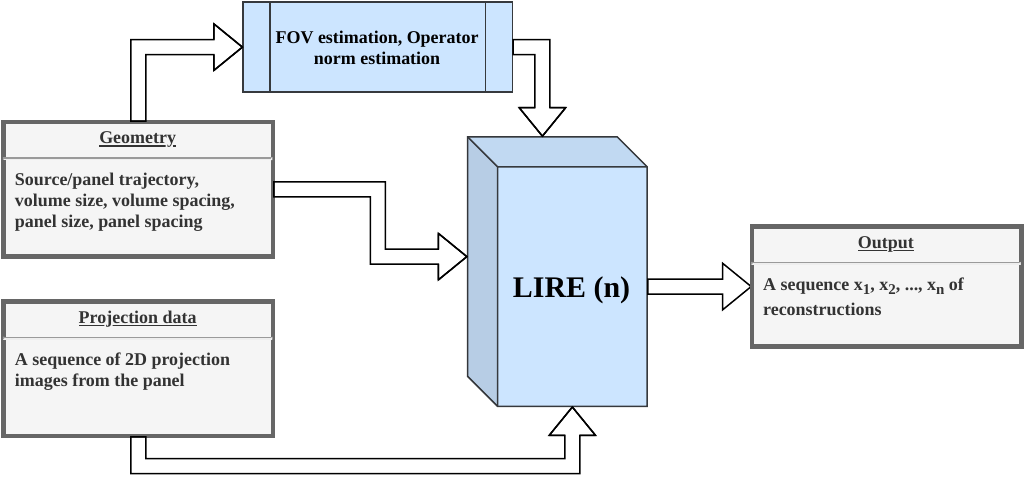}
    \caption{LIRE flowchart.}
    \label{fig:graph-abs}
\end{figure*}

LIRE is a data-driven algorithm, where a learned iterative scheme is unrolled and the parameters of this scheme are jointly optimized to minimize expected reconstruction loss over the training dataset; we summarize this in a flowchart in Figure \ref{fig:graph-abs}. The choice of a particular scheme will, naturally, affect both the performance and the required resources such as GPU memory to train such a scheme. 

\vspace{20px}

\begin{figure*}[h]
    \centering
    \includegraphics[width=0.7\linewidth]{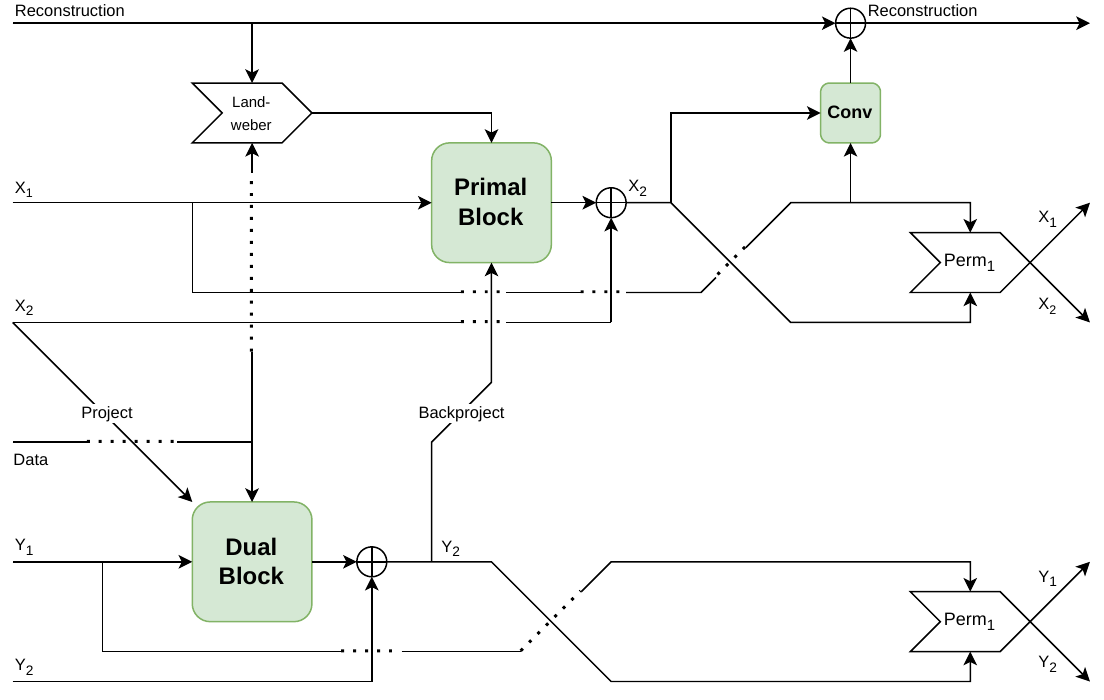}
    \caption{LIRE schematics.}
    \label{fig:lire-schm}
\end{figure*}

When designing LIRE, we took inspiration from Learned Primal-Dual (LPD) reconstruction algorithm\cite{Adler2017b}. The main disadvantage of LPD, however, is that it does not scale well to 3D reconstruction problems such as Cone Beam CT. We drastically reduce memory footprint of the LIRE algorithm compared to vanilla LPD and at the same time improve its expressive power by using more complex primal and dual blocks. In order to reduce the memory footprint, we designed LIRE around two main principles: invertibility for network as a whole and patch-wise computations for local operations. We briefly describe these two concepts below. Furthermore, for the additional full resolution experiment on high resolution data, we implemented a CPU-GPU memory streaming mechanism, which would keep entire primal/dual vectors in CPU memory and only send the channels required for computing the primal/dual updates to the GPU.

In an invertible residual network layer\cite{revnet2017}, the input tensor $z$ is split into tensors $z_1, z_2$ along the channel dimension. The output $w$ of the layer is then defined by combining $z_1$ and $z_2 + \Lambda(z_1)$ back along the channel dimension, where $\Lambda$ is a Convolutional Neural Network. Since the input $z$ can be uniquely restored\footnote{Up to numerical errors, which are typically negligible in practice for small neural networks.} from the output $w$, it is not essential to store intermediate features of $\Lambda$ prior to the actual computation of gradients for $\Lambda$. The main observation behind patch-wise computations is that for neural networks, which are composed solely of local operators such as valid convolutions, activation functions, upsampling and downsampling layers, it is possible to partition the input tensor $z$ into patches $z_i, i=1,\dots,k$ along the spatial dimensions and compute the output patch-by-patch. In general, for every $i$ it is also necessary to enlarge the patch $z_i$ by an adding all tensor elements $\partial z_i \subset z$ within a certain distance to $z_i$ in order to account for the receptive field of the network when computing features for locations inside $z_i$. For a special case of image classification CNNs that do not involve U-net type architectures or non-local operations such a patch-wise computation strategy was used previously\cite{pinckaers2019}. 

The reconstruction algorithm is given by the function \texttt{RECONSTRUCT}($y, \mathcal P, \mathcal P^*, \theta, V$) in Algorithm~\ref{alg:liremain}, while a simplified schematic depiction for the first reconstruction step is additionally provided in Figure~\ref{fig:lire-schm}. Here $y$ is the log-transformed and scaled projection data, $\mathcal P$ and $\mathcal P^*$ are the normalized projection and backprojection operators respectively, $\theta$ is a parameter vector and $V$ is an auxiliary single-channel image space tensor with the same dimensions as the reconstructed volume which we will define later in Section~\ref{ss.liretrain}. Parameters $\theta$ are partitioned into 4 parameter groups, where $\{ \theta_i^p \}_{i=1}^8$ are the primal block parameters, $\{ \theta_i^d \}_{i=1}^8$ are the dual block parameters, $\{ \theta_i^o \}_{i=1}^8$ are the output convolution parameters and $\{ \theta_i^m \}_{i=1}^8$ are the permutation parameters.

We clarify the notation first. We write $[z_1, z_2, \dots, z_k]^{\oplus}$ to denote the channel-wise concatenation of tensors $z_1, z_2, \dots, z_k$ which are assumed to have the same spatial and batch dimensions. Function $\text{\texttt{Splt}}(z)$ splits tensor $z$ with $2n$ channels into tensors $z_1, z_2$ which get the first $n$ feature maps of $z$ and the last $n$ feature maps of $z$ respectively. Function $\text{\texttt{Perm}}(z, {\theta_i^m})$ permutes tensor $z$ with $n$ channels along the channel dimension with the permutation $\theta_i^m \in \text{\texttt{Sym}}(n)$.

In LIRE we use 8 primal/dual iterations (8 primal and 8 dual blocks) with both primal and dual latent vectors having 8 channels. Backprojected data without FBP filtering is used to initialize the reconstruction $x_0$. The initial primal vector is defined by stacking 8 copies of $x_0$, and the initial dual vector is defined by stacking 8 copies of $y$. At the beginning of each iteration $i = 1,\dots,8$, we split the primal and the dual latent vectors along the channel dimension. First we update the dual latent vector in Line 10 of Alg.~\ref{alg:liremain} using dual block $\Gamma_{\theta_i^d}$ comprised of 3 layers of $3 \times 3 \times 3$ convolutions with 96, 96 and 4 filters respectively and LeakyReLU activation after the first and the second convolution layer. 
To update the primal block, we compute the Landweber term in Line 12 of Alg.~\ref{alg:liremain}, which plays a similar role as the gradient log-likelihood term in Recurrent Inference Machines\cite{lonning2019}. We update the primal latent vector in Line 13 of Alg.~\ref{alg:liremain} using primal block $\Lambda_{\theta_i^p}$. Primal block $\Lambda_{\theta_i^p}$ is a U-net with a single downsampling layer, $3 \times 3 \times 3$ valid convolutions with 96 filters in the first double-convolution block, 192 filters in the bottleneck and LeakyReLU activation after all but the last convolution layer. We use average pooling with $2 \times 2 \times 2$ kernel in the encoder and nearest upsampling in the decoder layer.
Primal and the dual updates are computed patch-wise, which is possible thanks to the locality of $\Gamma_{\theta_i^d}$ and $\Lambda_{\theta_i^p}$, during backward pass weight gradients obtained from patches are summed to obtain the global weight gradients. New primal and dual vectors are combined in Lines 14-15. Reconstruction $x_{i-1}$ is updated in Line 16, where $\text{\texttt{Conv3d}}$ is a $1 \times 1 \times 1$ convolution with parameters $\theta_i^o$, and we append the new reconstruction $x_i$ to the output list in Line 17. Finally, we permute the channels of primal and dual latent vectors using the same permutation $\theta_i^m$ in Lines 18-19. For every $i$, the permutation $\theta_i^m$ is some fixed permutation of $[1,2,\dots,8]$ which is randomly initialized during model initialization and stored as a model parameter; we require that $\theta_i^m$ mixes the first and the second half of $[1,2,\dots,8]$. The goal of permuting the channels randomly is to improve the flow of information between invertible blocks, which only update half of the primal or dual channels at once each time.

The algorithm was implemented as a `black box' C++/CUDA extension for PyTorch\cite{pytorch} in order to maximize speed and memory efficiency. Firstly, we implemented the projection and the backprojection operators for CBCT geometry as a CUDA extension for PyTorch. Since both operators are linear and the backprojection operator is the Hermitian adjoint of the projection operator, this is sufficient to enable gradient backpropagation. In the projector code, we followed the same approach as ASTRA Toolbox\cite{astra2016} and PYRO-NN\cite{syben2019} by using texture memory and trilinear interpolation when sampling attenuation values along the source-detector rays. Adjointness of the operators was tested by checking the definition of Hermitian adjoint
\begin{equation*}
\langle \mathcal P x, y \rangle = \langle x, \mathcal P^* y\rangle
\end{equation*}
for random positive test functions (=tensors) $x, y$. The LIRE network itself was then built as a C++/CUDA extension for PyTorch by implementing \textit{both} forward and backward passes since automatic differentiation is not available inside C++/CUDA extensions. PyTorch automatic differentiation was still used to compute the gradients of the loss for the output tensors $x_1, \dots, x_8 \in I$, but the subsequent computation of the gradients of the parameters $\theta$ was performed by LIRE in the backward pass. Correctness of the gradient computations for LIRE parameters was verified by computing numerical directional derivatives for random directions inside the parameter space and comparing this with the analytical directional derivatives computed using gradients from LIRE.

\subsection{LIRE training procedure}
\label{ss.liretrain}
\begin{table*}[t]
\caption{Test results on toy thorax CT, large FOV at 4 mm voxel pitch (best result in bold), mean $\pm$ std.dev.}
\label{tab:comparison-lung-toy}
\centering
\begin{tabular}{|l|l l|l|l|l|}
\hline
\multirow{2}{*}{Method} & \multicolumn{2}{c|}{Thorax CT} & \multirow{2}{*}{\# weights} \\
\cline{2-3}
  & PSNR & SSIM & \\
  \hline
DL-PWLS & $26.85 \pm 1.93$ & $0.74 \pm 0.07$ & 2456k \\
iLPD & $26.93 \pm 1.95$ & $0.74 \pm 0.07$ & 672k \\
LPD & $28.87 \pm 1.95$ & $0.82 \pm 0.06$ & 602k \\
$\partial$U-Net-16 & $28.04 \pm 1.95$ & $0.78 \pm 0.06$ & 6671k \\
$\partial$U-Net-32 & $28.39 \pm 1.94$ & $0.80 \pm 0.06$ & 26660k \\
LIRE-32(L) & $28.83 \pm 1.98$ & $0.81 \pm 0.06$ & 643k \\
LIRE-32 & $\mathbf{29.88 \pm 1.99}$ & $\mathbf{0.85 \pm 0.05}$ & 2857k \\
\hline
\end{tabular}
\end{table*}

\begin{table*}[t]
\caption{Test results on thorax CT and head \& neck CT at 2 mm voxel pitch (best result in bold), mean $\pm$ std.dev.}
\label{tab:comparison-lung}
\centering
\begin{tabular}{|l|l l|l l|l|l|l|}
\hline
\multirow{2}{*}{Method} & \multicolumn{2}{c|}{Thorax CT} & \multicolumn{2}{c|}{H\&N CT} & \multirow{2}{*}{\# weights} \\
\cline{2-5}
  & PSNR & SSIM & PSNR & SSIM & \\
\hline
FBP (small FOV) & $15.29 \pm 2.60$ & $0.57 \pm 0.09$ & $27.64 \pm 1.69$ & $0.72 \pm 0.03$ & - \\
TV (small FOV) & $27.37 \pm 2.64$ & $0.77 \pm 0.09$ & $33.90 \pm 1.05$ & $0.86 \pm 0.03$ & - \\
Uformer (small FOV) & $29.76 \pm 1.84$ & $0.72 \pm 0.07$ & $29.78 \pm 0.82$ & $0.72 \pm 0.04$ & 82251k \\
U-Net (small FOV) & $33.08 \pm 1.75$ & $0.81 \pm 0.05$ & $36.45 \pm 0.84$ & $0.88 \pm 0.01$ & 23341k \\
LIRE (small FOV) & $\mathbf{33.84 \pm 2.28}$ & $\mathbf{0.89 \pm 0.05}$ & $\mathbf{39.35 \pm 1.75}$ & $\mathbf{0.96 \pm 0.01}$ & 24497k \\
\hline
FBP (large FOV) & $20.05 \pm 2.30$ & $0.66 \pm 0.07$ & $22.39 \pm 0.44$ & $0.71 \pm 0.02$ & - \\
TV (large FOV) & $29.23 \pm 2.87$ & $0.79 \pm 0.09$ & $37.86 \pm 1.36$ & $0.94 \pm 0.02$  & - \\
Uformer (large FOV) & $31.62 \pm 2.44$ & $0.81 \pm 0.06$ & $29.90 \pm 0.87$ & $0.86 \pm 0.01$ & 82251k \\
U-Net (large FOV) & $34.29 \pm 2.71$ & $0.84 \pm 0.06$ & $37.06 \pm 1.21$ & $0.88 \pm 0.01$ & 23341k \\
LIRE (large FOV) & $\mathbf{35.14 \pm 2.69}$ & $\mathbf{0.91 \pm 0.05}$ & $\mathbf{41.21 \pm 1.41}$ & $\mathbf{0.97 \pm 0.01}$ & 24497k \\
\hline
\end{tabular}
\end{table*}

\begin{table*}[t]
\caption{Test results on thorax CT at 1 mm voxel pitch (best result in bold)}
\label{tab:comparison-lung-high}
\centering
\begin{tabular}{|l|l l|l|l|l|}
\hline
\multirow{2}{*}{Method} & \multicolumn{2}{c|}{Thorax CT} & \multirow{2}{*}{\# weights} & \multirow{2}{*}{Opt. steps}\\
\cline{2-3}
  & PSNR & SSIM &  & \\
\hline
U-Net (large FOV) & $33.773$ & $0.848$ & 23341k & 62400 (cold start) \\
LIRE (large FOV) & $\mathbf{35.784}$ & $\mathbf{0.881}$ & 24497k & 1560 (warm start) \\
\hline
\end{tabular}
\end{table*}
We provide the training procedure for LIRE in Algorithm~\ref{alg:liretrain}. The training is supervised, and the training set of CT volumes is denoted by $\mathcal D_{\text{train}}$. We elaborate on the training procedure below.

A CT volume is repeatedly sampled from the training dataset in Line 9 of Alg.~\ref{alg:liretrain}. During the sampling, augmentations that flip patient left-right and top-bottom are randomly applied, both with probability $50 \%$. We sample a random offset for the rotation center w.r.t. the center of the CT volume from an isotropic Gaussian distribution with $0$ mean and a standard deviation of $100$ mm in Line 10. Choosing a random offset can be viewed as an additional type of augmentation, furthermore, in practice the isocenter in radiotherapy will be located close to a tumor. We define projection and backprojection operators for the CBCT projection geometry with given volume spacing and center offset in Line 12, and in Line 13 we compute normalized versions of these operators. The operator norm is estimated numerically using power method with three iterations\cite{boyd1974}. Synthetic noisy projection data is computed in Line 14 (see \eqref{eq.noisemodel}). This noisy projection data is log-transformed and scaled in Line 15. In general, for a realistic CBCT geometry the field of view does not necessarily contain scanned object completely. When comparing reconstruction metrics it is also important to compute these metrics inside an appropriately defined field of view only, since having a large part of the volume set to $0$ outside the corresponding field of view would yield over-optimistic reconstruction metrics. We define the full field of view tensor $V_f$ and the partial field of view tensor $V_p$ in Lines 16 and 17 respectively, both of these are scalar tensors having same dimensions as the volume that we want to reconstruct. For the projection geometry with small FOV setting, the full field of view tensor is constructed as
\begin{equation*}
V_f(p) = \begin{cases} 
      1 & p \textrm{ is seen from all projection angles} \\
      0 & \textrm{otherwise,}
   \end{cases}
\end{equation*}
while for the projection geometry with large FOV setting the full field of view tensor is constructed as
\begin{equation*}
V_f(p) = \begin{cases} 
      1 & p \textrm{ is seen from all projection angles} \\
      0.5 & p \textrm{ is seen from half of the proj. angles} \\
      0 & \textrm{otherwise.}
   \end{cases}
\end{equation*}
We chose to use different values ($1.0$ and $0.5$) above to mark the voxels seen from all the projection angles and the voxels which are seen from only half of the angles, however, we expect that the exact numerical values used in these masks are not important. For both small and large field of view settings, the partial field of view is defined as
\begin{equation*}
V_p(p) = \begin{cases} 
      1 & p \textrm{ is seen from at least one angle} \\
      0 & \textrm{otherwise.}
   \end{cases}
\end{equation*}
In particular, this definition of $V_p$ implies that in the central axial plane all voxels are marked as `partially visible'. In Line 18, we define $V_s$ as the tensor which equals $1$ on the set of all voxels $p$ s.t. $V_p(p) > 0, V_f(p) = 0$ and zero elsewhere. In Line 19, we call the main reconstruction procedure, providing log-transformed normalized projection data, normalized versions of projection and backprojection operators, the collection of weights $\theta$ and the auxiliary tensor $V_f$. $V_f$ helps the network to deal with the non-homogeneouos nature of the reconstruction artifacts. 

The reconstruction algorithm returns a list $I = [z_1, z_2, \dots, z_8]$ of reconstructions, which are obtained after performing $1, 2, \dots, 8$ reconstruction steps respectively. We sum the reconstruction losses over all $z \in I$ in Line 22. Loss computation takes place in the $\texttt{LOSS}$ function in Alg.~\ref{alg:liretrain}. We sum losses over the full field of view region, where $V_f > 0$, and the partial field of view region, where $V_s > 0$. We compute the loss for partial field of view to ensure that the network can provide at least an approximate reconstruction in this region. A linear combination of $L^1$ loss and Structural Similarity Loss is computed for both regions. We used $\alpha_1 = 0.5$ for both field of view settings. $\alpha_2$ was set to $0.1$ initially and then reduced to $0.01$ after first learning rate decay step.

We trained two versions of LIRE for the main experiments, one for the small FOV setting and one for the large FOV setting. LIRE was trained to reconstruct complete volumes. For the internal patch-based computations inside LIRE we set the patch size to $128 \times 128 \times 128$, resulting in roughly $30$ GB VRAM usage per single volume. Reducing the patch size to $32 \times 32 \times 32$ decreased the usage to roughly $20$ GB VRAM per single volume. Eight NVIDIA Quadro RTX 8000 GPUs with 48 GB VRAM were used for training LIRE in distributed data parallel mode to achieve the batch size of 8. We used Adam optimizer\cite{kingma2014} with an initial learning rate of $0.0001$ and a plateau scheduler with linear warm-up and 10 epoch patience. At the end of each epoch models were evaluated, the best model was picked for testing. Training was stopped when we did not observe improvement for more than 15 epochs. For the additional study on high-resolution data, we performed a warm start from LIRE trained on downsampled lung CT scans with large FOV setting. Two NVIDIA A100 Tensor Core GPUs with 80 GB VRAM inside a virtual machine on NVIDIA TryEGX Platform were used. We employed Adam optimizer with an initial learning rate of $0.000025$ and linear warm-up, the model was fine-tuned for 12 epochs without any learning rate decay. $L^1$ reconstruction loss was used during fine-tuning, due to higher memory costs associated with SSIM loss.

\subsection{LIRE memory analysis}
\label{ss.liremem}
It should be noted that the memory costs associated with storing the primal/dual vectors and the inputs to the primal/dual blocks are noticeably higher than the costs of computations inside the blocks thanks to the tiling mechanism. This does, in particular, result in lower memory usage in comparison with iLPD, which allows us to use more complex architectures inside the primal/dual blocks. In the additional high resolution study we reduce the memory costs associated with storing primal/dual vectors by storing primal/dual latent vectors in the CPU memory and loading only the channels which are currently needed into the GPU memory. This enables us to train our algorithm for reconstruction of volumes with $512^3$ voxels at $1$ mm resolution with $512 \times 512$ detector.

\section{Results}
\subsection{Inference times and evaluation metrics}
\label{s.evaldet}
For the main experiments on downsampled data, it takes U-net+FBP approximately 10 seconds to reconstruct volumes for both geometries, while Uformer+FBP takes approximately 50 seconds. PDHG takes 14 minutes to reconstruct a small FOV volume and 18 minutes to reconstruct a large FOV volume\footnote{It should be noted that the ODL implementation of PDHG performs a lot of data transfer between CPU and GPU, which hurts the performance.}.

During the inference on the downsampled data with 2 mm voxel pitch and $256 \times 256$ detector panel, it takes LIRE approximately 104 seconds to reconstruct a single volume on a single Quadro RTX 8000 for the small FOV setting and approximately 115 seconds to reconstruct a volume for the large FOV setting. On high-resolution data with 1 mm voxel pitch, $512 \times 512$ detector panel and large FOV setting, it takes LIRE approximately 15 minutes to reconstruct a single volume on Quadro RTX 8000, and 6 minutes if A100 is used instead. Faster inference on A100 can be attributed to higher memory bandwidth and other hardware architectural improvements.

Evaluation of LIRE and the baseline methods was performed using PSNR and SSIM metrics restricted to the full field of view region, where $V_f > 0$. This corresponds to the region visible from all projection angles in the case of small FOV setting, and to the region visible from at least half of the projecion angles in the case of large FOV setting.

\subsection{Toy data}
\label{s.tresults}
In Table \ref{tab:comparison-lung-toy} we summarize the results for the toy test set of thorax CT scans, where we compare our method against other learned iterative schemes. Firstly, we note that LIRE-32 outperforms alternative baselines. LIRE-32(L), which is a simplified version of LIRE-32 architecture with a stack of 3 convolutional layers with 32 filters inside the primal blocks instead of U-Nets,  performs on par with LPD with very similar filter count, showing that invertible architecture of the primal/dual blocks is not a limiting factor for the reconstruction quality. iLPD is also outperformed by LIRE-32(L) despite similar parameter counts, which could be attributed to the channel mixing strategy between invertible blocks of LIRE which does not appear in iLPD that internally relies on MemCNN\cite{vandeLeemput2019MemCNN}. iLPD performs slightly worse than LPD in this experiment despite similar parameter count, which is in agreement with the previous results\cite{rudzusika2021}. Since LIRE-32 outperforms LIRE-32(L), we can also conclude that using U-Nets inside the primal blocks leads to a performance gain. At the same time, the multiscale $\partial$U-Net architectures $\partial$U-Net-16 and $\partial$U-Net-32 are outperformed by LPD and LIRE despite having much higher filter counts. It should be noted that $\partial$U-Nets use a U-Net variant with 4 downsampling layers which merges reconstructions obtained at different resolutions, while LIRE-32 relies on U-Nets with a single downsampling layer inside each primal block. 

\subsection{Main experiments: full field of view}
\label{s.results}

\begin{figure*}[!ht]
    \centering
    \subfloat[]{\includegraphics[width=0.25\linewidth]{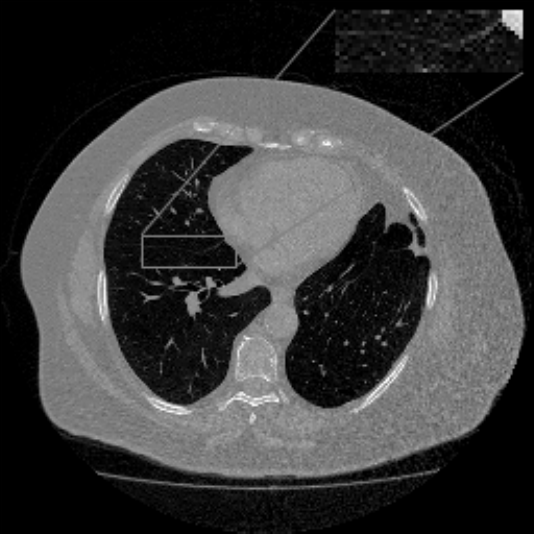}
    }
    \hfil
    \subfloat[]{\includegraphics[ width=0.25\linewidth]{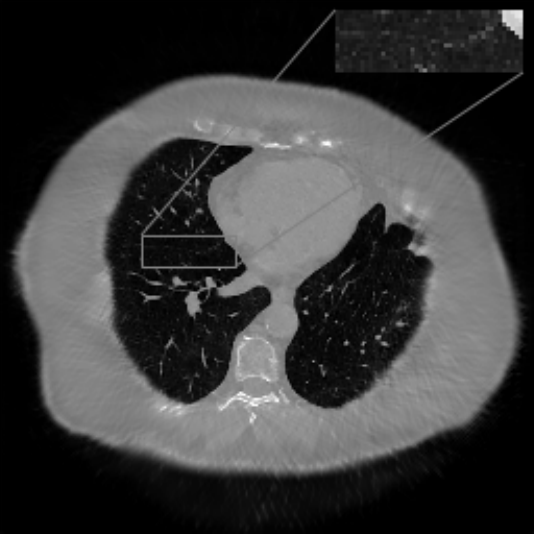}
    }
    \hfil
    \subfloat[]{\includegraphics[ width=0.25\linewidth]{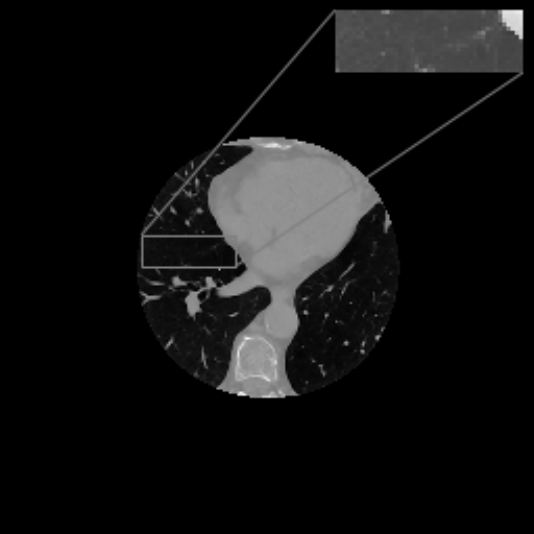}
    }
    \hfil
    \subfloat[]{\includegraphics[width=0.25\linewidth]{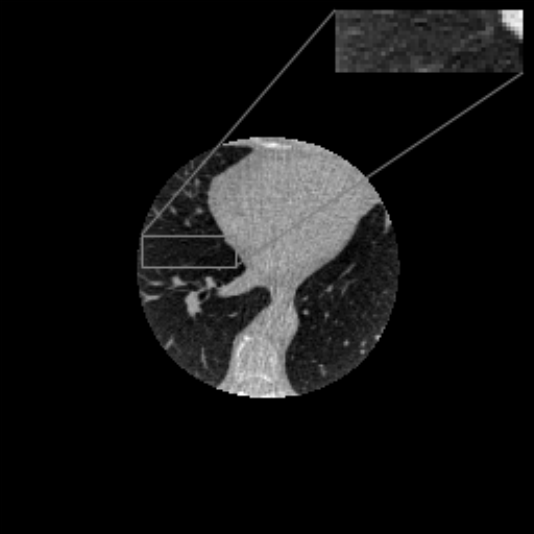}
    }
    \caption{(a) Axial slice of Thorax CT with HU range=(-1000, 800) and (-1350,150) for ROI, (b) LIRE/small FOV, (c) U-net/small FOV, and (d) PDHG/small FOV.}
    \label{fig:axial-small}
\end{figure*}

\begin{figure*}[!ht]
    \centering
    \subfloat[]{\includegraphics[width=0.25\linewidth]{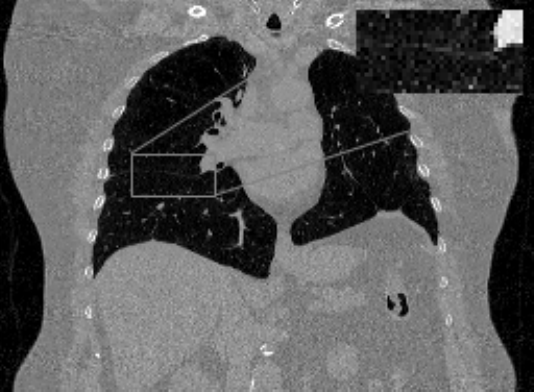}
    }
    \hfil
    \subfloat[]{\includegraphics[ width=0.25\linewidth]{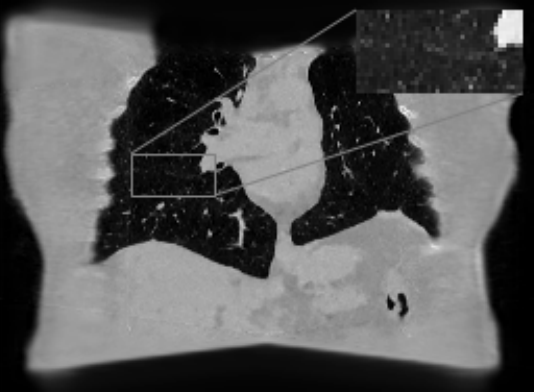}
    }
    \hfil
    \subfloat[]{\includegraphics[ width=0.25\linewidth]{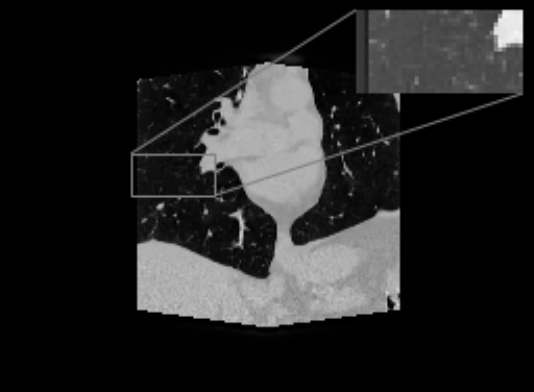}
    }
    \hfil
    \subfloat[]{\includegraphics[width=0.25\linewidth]{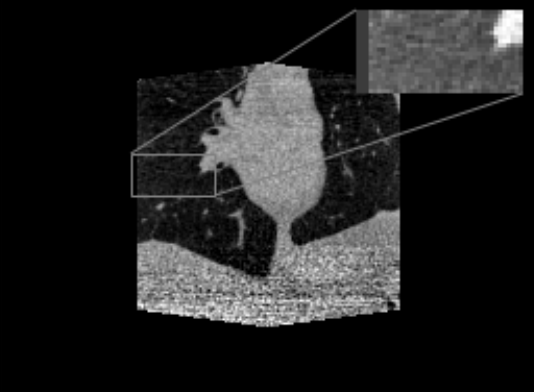}
    }
    \caption{(a) Coronal slice of Thorax CT with HU range=(-1000, 800) and (-1350,150) for ROI, (b) LIRE/small FOV, (c) U-net/small FOV, and (d) PDHG/small FOV.}
    \label{fig:coronal-small}
\end{figure*}

\begin{figure*}[!ht]
    \centering
    \subfloat[]{\includegraphics[width=0.25\linewidth]{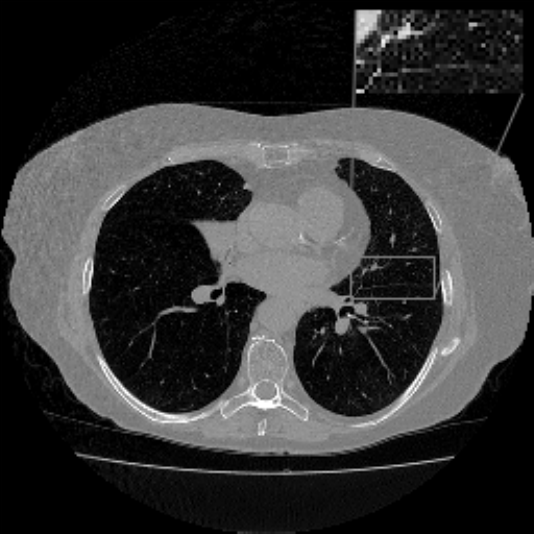}
    }
    \hfil
    \subfloat[]{\includegraphics[ width=0.25\linewidth]{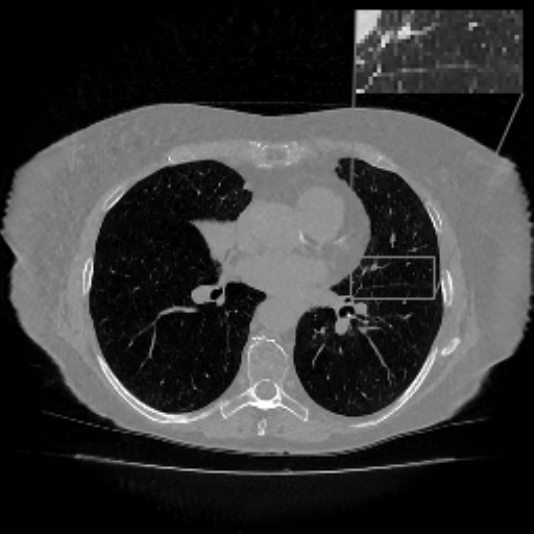}
    }
    \hfil
    \subfloat[]{\includegraphics[ width=0.25\linewidth]{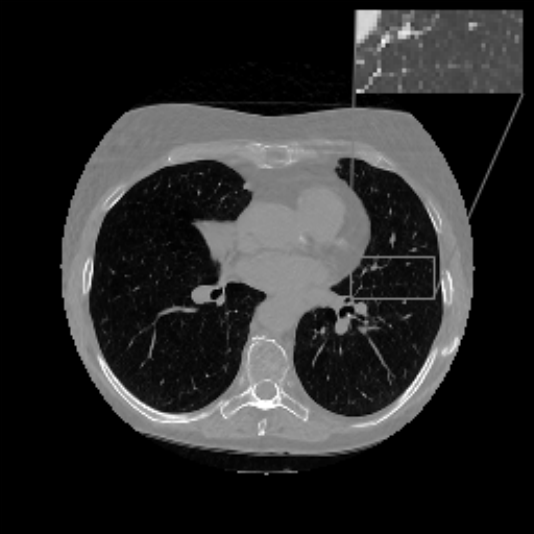}
    }
    \hfil
    \subfloat[]{\includegraphics[ width=0.25\linewidth]{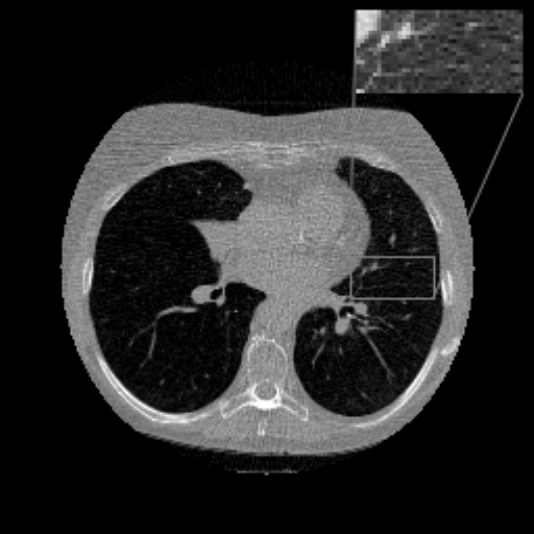}
    }
    \caption{(a) Axial slice of Thorax CT, HU range=(-1000, 800) and (-1350,150) for ROI, (b) LIRE/large FOV, (c) U-net/large FOV, and (d) PDHG/large FOV.}
    \label{fig:axial-large}
\end{figure*}

\begin{figure*}[!ht]
    \centering
    \subfloat[]{\includegraphics[width=0.25\linewidth]{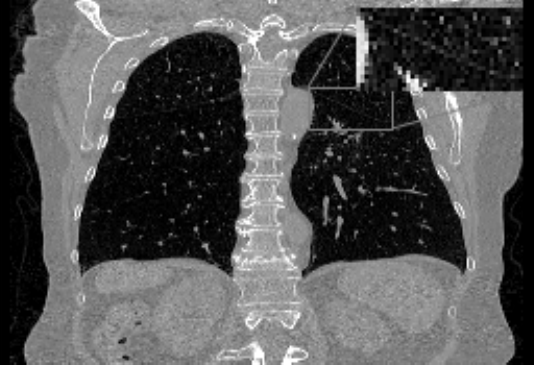}
    }
    \hfil
    \subfloat[]{\includegraphics[ width=0.25\linewidth]{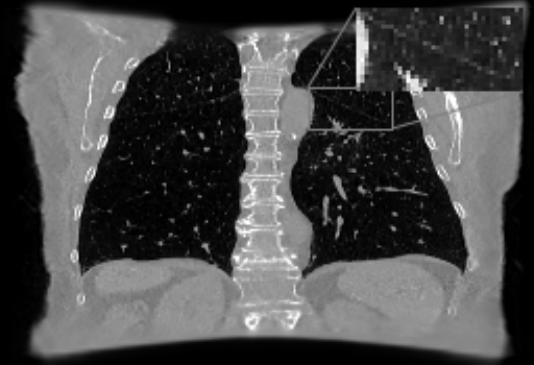}
    }
    \hfil
    \subfloat[]{\includegraphics[ width=0.25\linewidth]{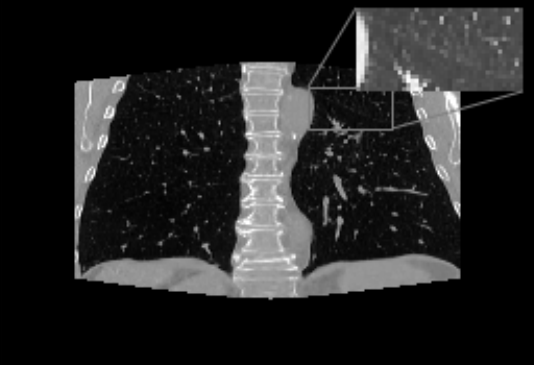}
    }
    \hfil
    \subfloat[]{\includegraphics[ width=0.25\linewidth]{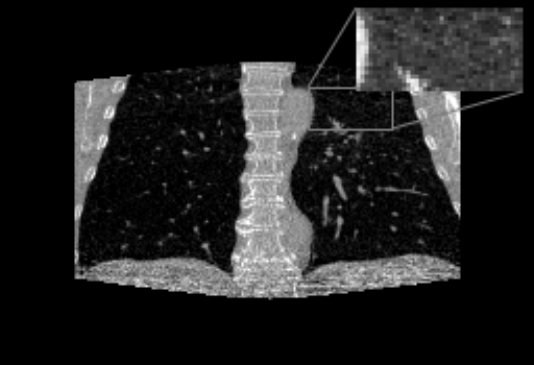}
    }
    \caption{(a) Coronal slice of Thorax CT, HU range=(-1000, 800) and (-1350,150) for ROI, (b) LIRE/large FOV, (c) U-net/large FOV, and (d) PDHG/large FOV.}
    \label{fig:coronal-large}
\end{figure*}

\begin{figure*}[!ht]
    \centering
    \subfloat[]{\includegraphics[width=0.25\linewidth]{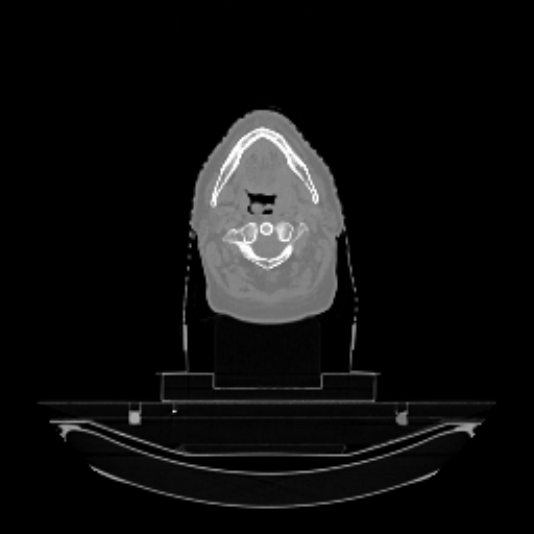}
    }
    \hfil
    \subfloat[]{\includegraphics[ width=0.25\linewidth]{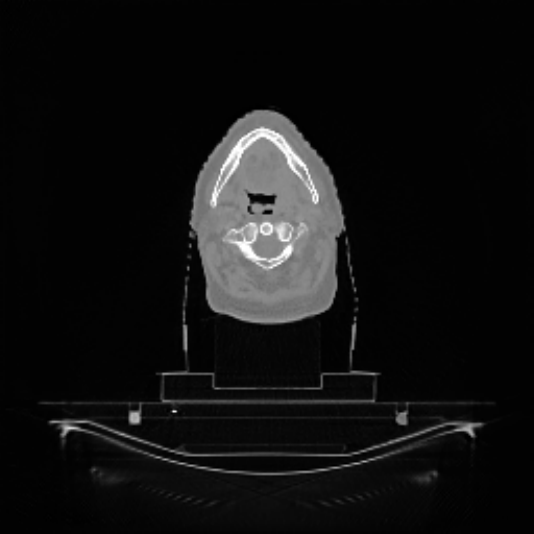}
    }
    \hfil
    \subfloat[]{\includegraphics[ width=0.25\linewidth]{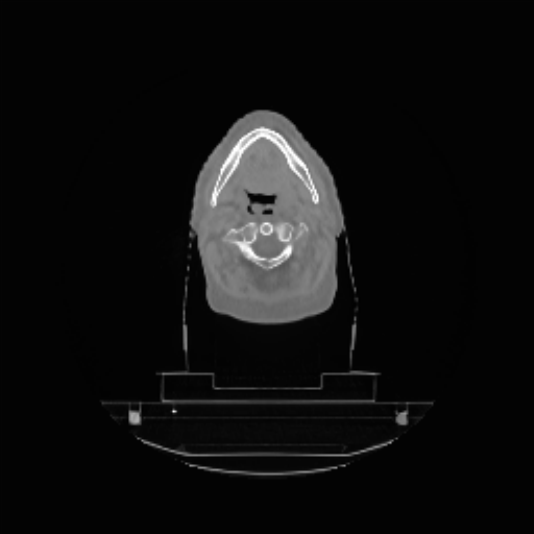}
    }
    \hfil
    \subfloat[]{\includegraphics[ width=0.25\linewidth]{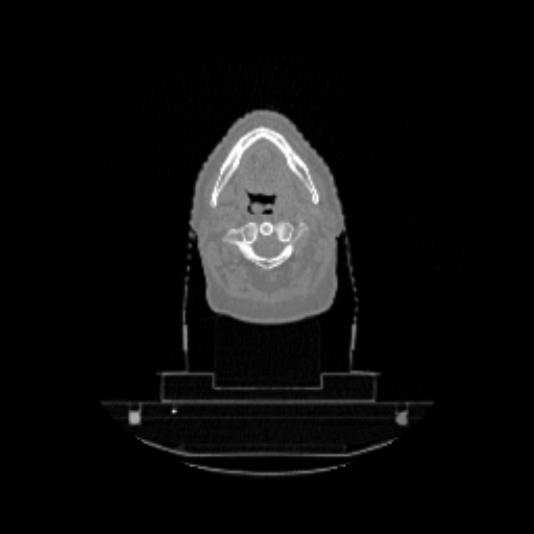}
    }
    \caption{(a) Axial slice of Head \& Neck CT with HU range=(-1000, 1000), (b) LIRE/large FOV, (c) U-net/large FOV, and (d) PDHG/large FOV.}
    \label{fig:axial-large-ext}
\end{figure*}

\begin{figure*}[!ht]
    \centering
    \subfloat[]{\includegraphics[width=0.25\linewidth]{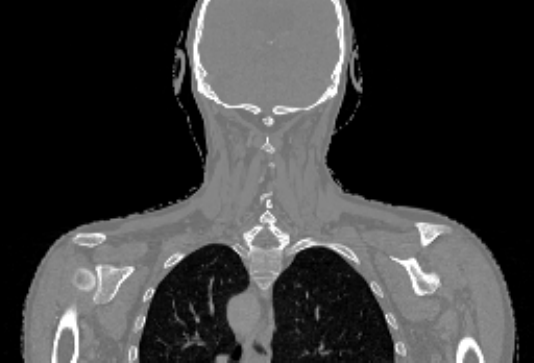}
    }
    \hfil
    \subfloat[]{\includegraphics[ width=0.25\linewidth]{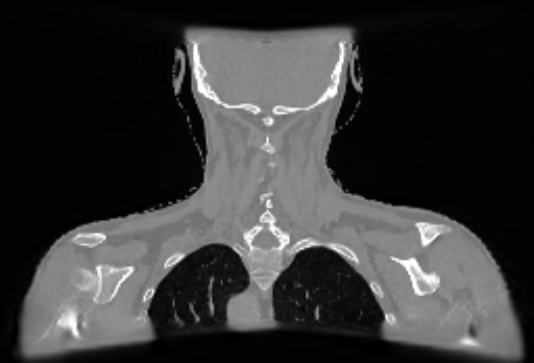}
    }
    \hfil
    \subfloat[]{\includegraphics[ width=0.25\linewidth]{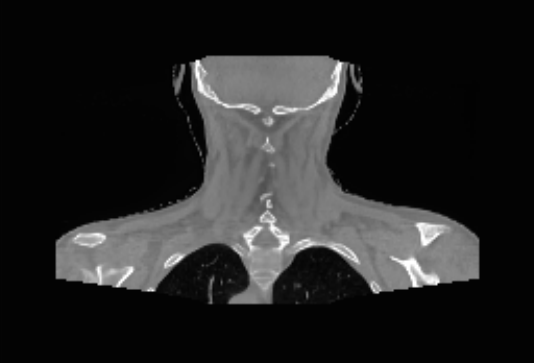}
    }
    \hfil
    \subfloat[]{\includegraphics[ width=0.25\linewidth]{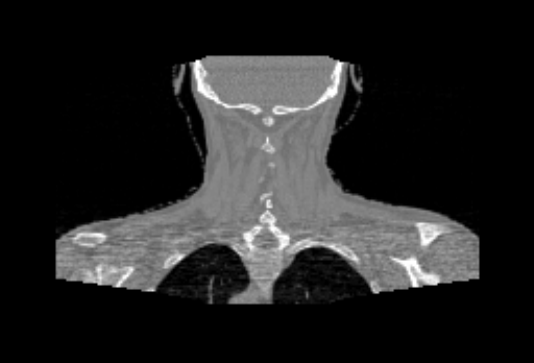}
    }
    \caption{(a) Axial slice of Head \& Neck CT with HU range=(-1000, 1000), (b) LIRE/large FOV, (c) U-net/large FOV, and (d) PDHG/large FOV.}
    \label{fig:coronal-large-ext}
\end{figure*}

In Table \ref{tab:comparison-lung} we summarize the results for the test set of thorax CT scans and the out-of-distribution test set of head \& neck CT scans. We provide thorax CT axial and coronal slices for small FOV in Figures \ref{fig:axial-small} and \ref{fig:coronal-small}, thorax CT axial and coronal slices for large FOV in Figures \ref{fig:axial-large} and \ref{fig:coronal-large}. Head \& neck CT axial and coronal slices for large FOV are given in Figures \ref{fig:axial-large-ext} and \ref{fig:coronal-large-ext}. The slices were taken from randomly chosen test volumes. We see that our method outperforms the classical and deep learning baselines in all cases, including the out-of-distribution test set. Uformer baseline, being a two-dimensional post-processing method, is outperformed by a three-dimensional U-net despite having more parameters. Compared to U-net+FBP, there is a notable is the improvement in SSIM, ranging from $+0.07$ to $+0.08$ on thorax CT data. PSNR improvement over the U-net is $+0.8$ dB for small FOV and $+0.9$ dB for large FOV. On the out-of-distribution test set, we observe better generalization of LIRE compared to U-net+FBP in the form of an increased PSNR and SSIM gap between LIRE and U-net, even though LIRE has a slightly higher parameter count, allowing to suggest that primal-dual schemes with shallow U-nets generalize better than a single deep U-net. Visual inspection of thorax CT slices shows better visibility of lung fissures in LIRE reconstructions compared to the baselines. In head \& neck CT slices, we observe that U-net loses spatial resolution and introduces a strong `shadow' in the neck region. LIRE yields best reconstructions on the head \& neck CT set due to better handling of photon noise compared to the iterative method, but in the low-noise neck region we observe that the methods are quite close in visual image quality. 

\begin{figure*}[!ht]
    \centering
    \subfloat[]{\includegraphics[width=0.33\linewidth]{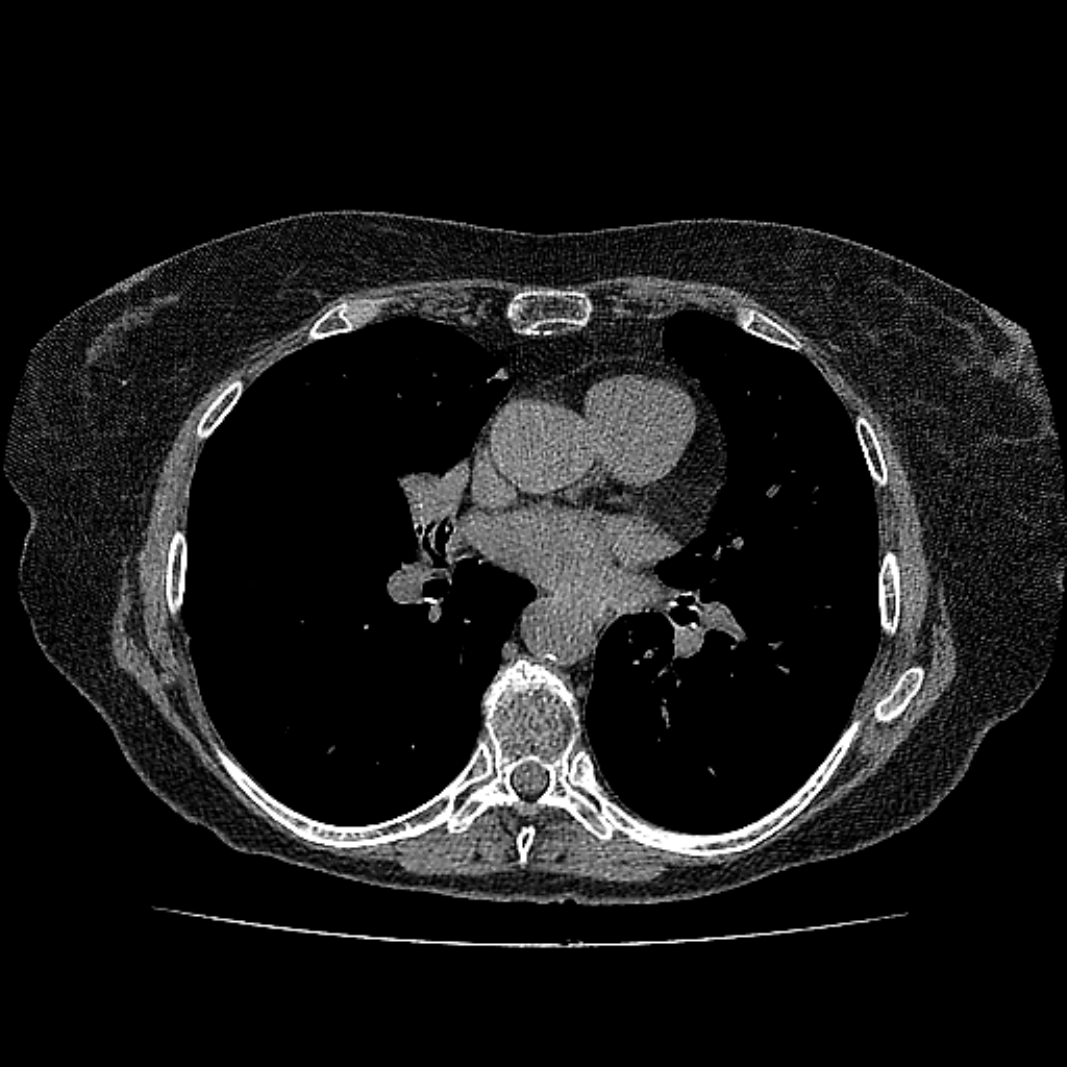}
    }
    \hfil
    \subfloat[]{\includegraphics[ width=0.33\linewidth]{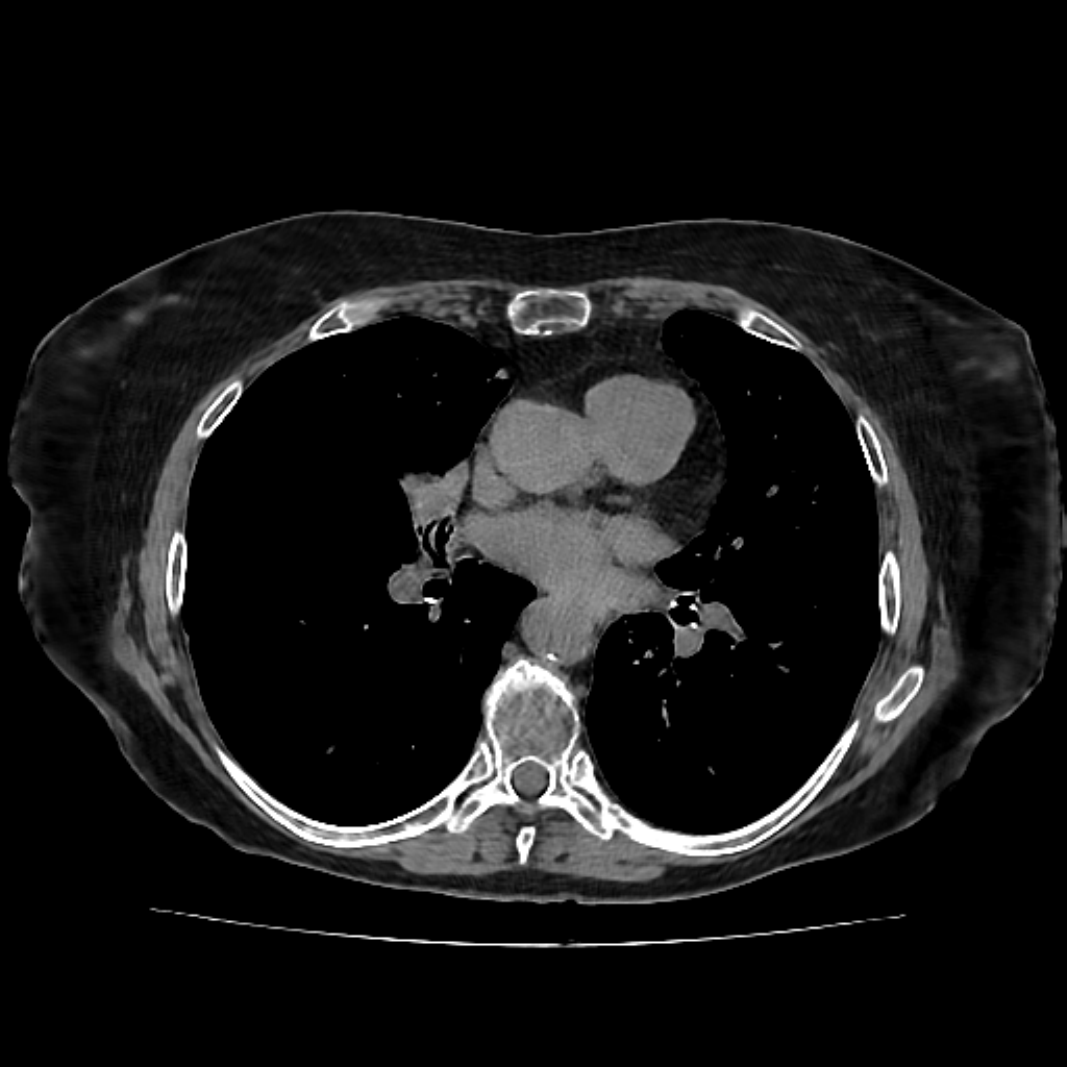}
    }
    \hfil
    \subfloat[]{\includegraphics[ width=0.33\linewidth]{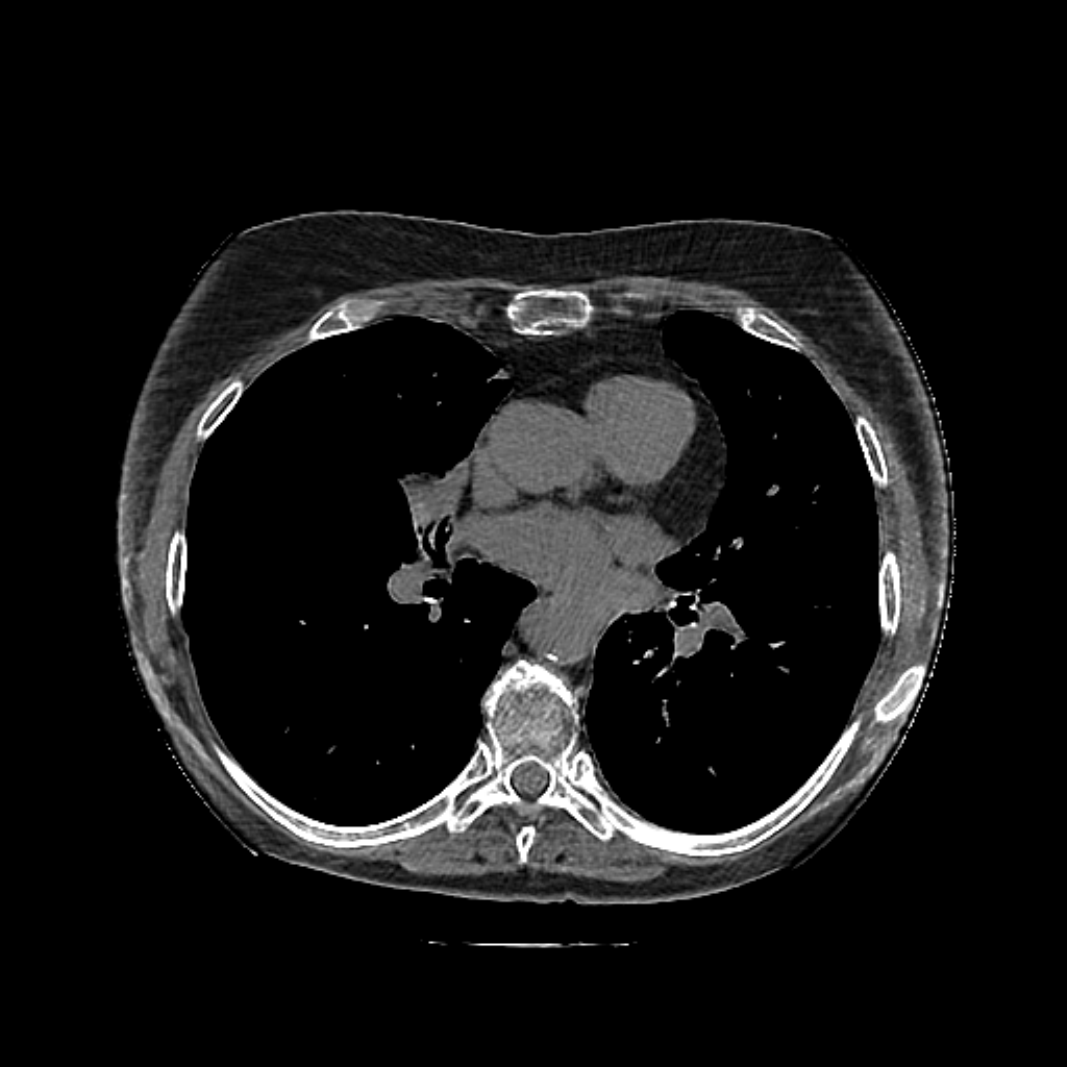}
    }
    \caption{(a) Axial slice of Thorax CT with HU range=(-150, 250) at 1mm resolution, (b) LIRE/large FOV, (c) U-net/large FOV.}
    \label{fig:axial-large-soft}
\end{figure*}

\begin{figure*}[!ht]
    \centering
    \subfloat[]{\includegraphics[width=0.33\linewidth]{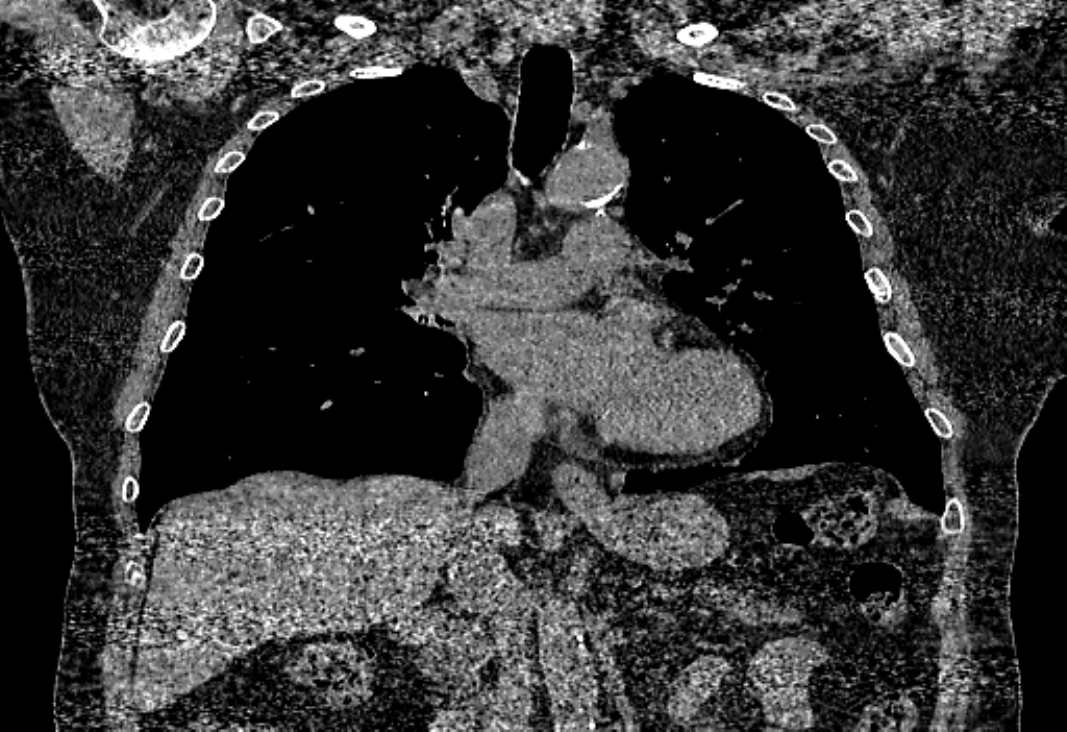}
    }
    \hfil
    \subfloat[]{\includegraphics[ width=0.33\linewidth]{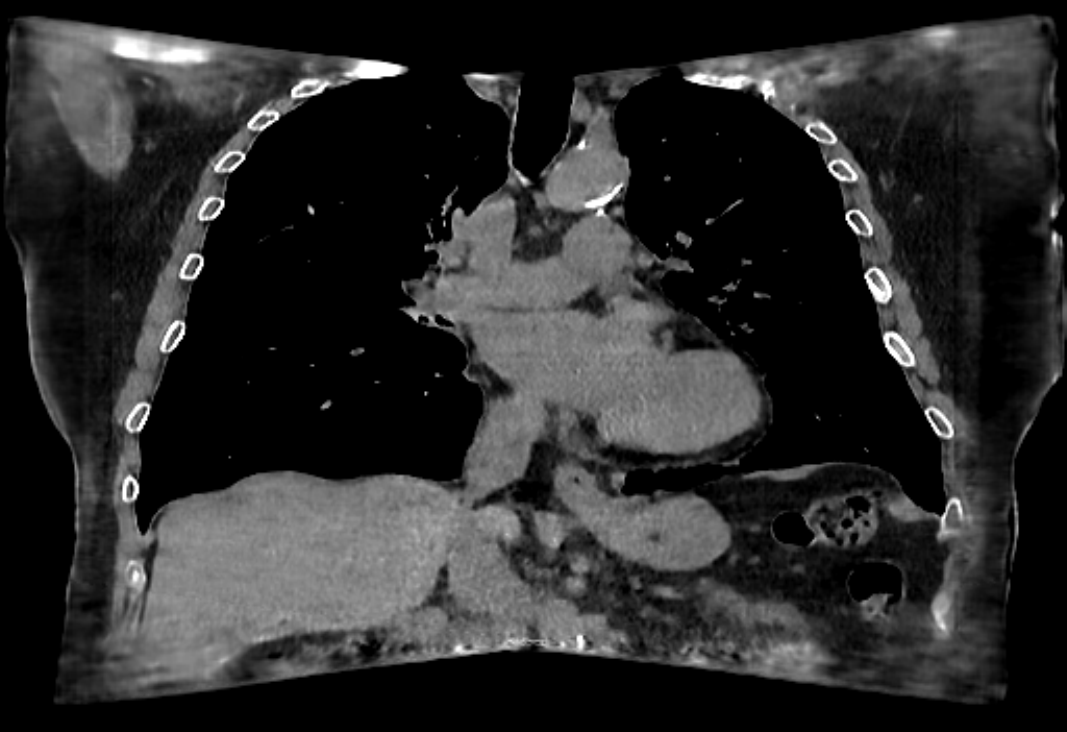}
    }
    \hfil
    \subfloat[]{\includegraphics[ width=0.33\linewidth]{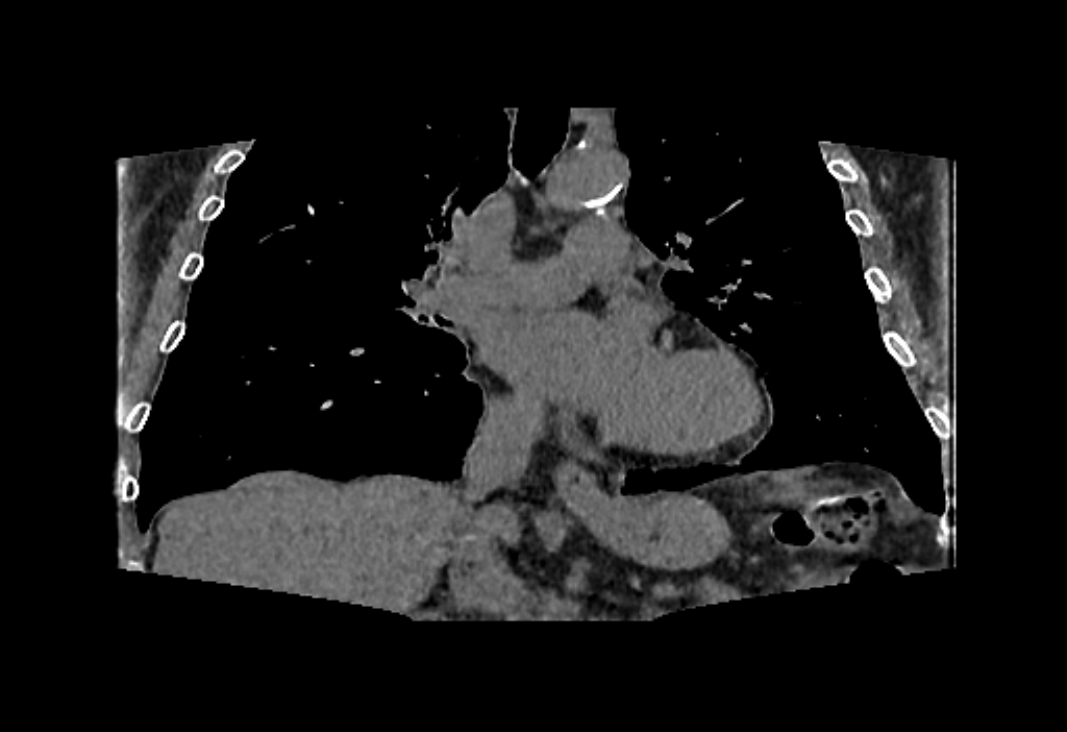}
    }
    \caption{(a) Coronal slice of Thorax CT with HU range=(-150, 250) at 1mm resolution, (b) LIRE/large FOV, (c) U-net/large FOV.}
    \label{fig:coronal-large-soft}
\end{figure*}

The results of the high-resolution experiment for the test set of thorax CT scans are summarized in Table \ref{tab:comparison-lung-high}, and we provide samples in Figures \ref{fig:axial-large-soft} and \ref{fig:coronal-large-soft}, where we used a different window level to highlight the quality of soft tissue reconstruction. We only provide comparison with the U-net+FBP, since it is the best performing baseline method on the downsampled data. We provide high-resolution thorax CT slices for large FOV as a video supplement. In these high resolution experiments, LIRE still outperforms U-net+FBP. The margin of improvement given by LIRE is increased, possibly due to the fact that the U-net has (spatially) smaller receptive field when working at higher resolution and, unlike LIRE, it cannot filter the noise in projection domain.

\subsection{Main experiments: partial field of view}
\label{s.fov}
Compared to U-net+FBP, another notable improvement is the much larger field-of-view in LIRE, since LIRE is not constrained by the data sufficiency region of FBP. 

We measured the performance of LIRE and PDHG on the test set of thorax CT data for the small FOV setting in the region where $V_f = 0, V_p = 1$, consisting of the voxels in the partial field of view which do not belong to the full field of view. This way we obtained mean PSNR of $16.938$ and mean SSIM of $0.233$ for PDHG, whereas for LIRE mean PSNR was $26.75$ and mean SSIM was $0.77$.

\section{Discussion}\label{sec:discussion}
\label{s.discussion}
We have presented LIRE, a practical algorithm for deep leaning-based CBCT reconstruction with clinically-relevant resolution and projection count using a learned primal-dual scheme that can be trained end-to-end on currently available GPUs with 24 GB VRAM. We have shown that our method outperforms the classical and deep learning baselines on the test set of thorax CT scans and the out-of-distribution test set of head \& neck CT scans, where we additionally observe better generalization of our method compared to the U-net baseline. In particular, the photon noise in highly attenuated areas is handled very well, which indicates that LIRE can potentially help to lower the dose of CBCT scans. For the small field of view setting, our method is able to reconstruct certain anatomy details outside the full field of view better than the iterative baseline, which can be interesting for applications in radiotherapy, e.g., by allowing for a better registration of the planning CT scan to the CBCT reconstruction. 

This work has certain limitations. Firstly, we do not take scatter artifacts into account. Feasibility of supervised scatter correction with deep learning was demonstrated by e.g. Deep Scatter Estimation method\cite{dse_spie}, and such method can be in principle combined with our learned primal-dual scheme and trained end-to-end. Secondly, we do not correct for possible motion artifacts in thorax CBCT due to breathing. Thirdly, our metrics do not directly imply suitability of our method for radiotherapy planning; a proper Monte Carlo dose simulation would be required to test that.

\section{Acknowledgment}
\label{s.acks}
We would like to thank NVIDIA Corporation for providing us with the access to A100 virtual machine instances and for supporting us throughout these experiments. In particular, we would like to thank Joe Cullen from NVIDIA for enabling this collaboration.

\section*{Conflict of Interest Statement}

The authors have no relevant conflicts of interest to disclose.






\end{document}